\documentclass[aps,floats,twocolumn,superscriptaddress,floatfix,nofootinbib]{revtex4}
\usepackage{amsfonts,amssymb,bm}
\usepackage{amsmath}
\usepackage{amscd}
\usepackage{bbm}
\usepackage{xy}
\usepackage{amssymb}
\usepackage{graphicx}
\usepackage{theorem}
\usepackage{makeidx}
\usepackage{amsmath}
\usepackage{epic}
\usepackage{epsfig}
\usepackage{times}

\newcommand{\ue}{\mathrm{e}}

\def\rset{{\rm I\kern -0.2em R}}
\def\un{\hbox{{1\kern -0.25em\raise 0.4ex\hbox{{\scriptsize $|$}}}}}

\def\cset{eq_burg_force\hbox{{C\kern -0.55em\raise 0.5ex\hbox{{\tiny $|$}}}}}
\def\nset{\hbox{{I\kern -0.18em N}}}

\def\twop{\boldsymbol{\mathfrak{p}}}
\def\twoq{\boldsymbol{\mathfrak{q}}}

\begin{document}

\title[On complex singularities]{On complex singularities of the 2D Euler equation at short times}
\author{W.~Pauls}
\affiliation{Max Planck Institute for Dynamics and Self-Organization, G\"ottingen, Germany}
\date{\today}
\begin{abstract}
We present a study of complex singularities of a two-parameter family of solutions for the 
two-dimensional Euler equation with periodic boundary conditions and initial conditions $\psi _0 (z_1 ,
z_2 ) = \hat{F} (\twop ) \cos \twop \cdot {\bm z} + \hat{F} (\twoq ) \cos \twoq \cdot {\bm z} $ in the short-time asymptotic r\'egime. 
As has been shown numerically in W.~Pauls {\it et al.}, Physica D{\bf 219}, 40--59 (2006), the type of the singularities depends on the angle $\phi $ between the modes $\twop $ and $\twoq $. Here we show for the two particular cases of $\phi $ going to zero and to $\pi $ that the type of the singularities can be determined very accurately, being characterised by 
$\alpha = 5/2 $ and $\alpha = 3$ respectively. In these two cases we are also 
able to determine the subdominant corrections. 
Furthermore, we find that the geometry of the singularities in 
these two cases is completely different, the singular manifold being located "over" different 
points in the real domain.
\end{abstract}
\vspace*{5mm}
\maketitle


\section{Introduction}
\label{s:introduction}

One of the most important open problems in mathematical fluid dynamics is the question of well-posedness of the three-dimensional incompressible Navier--Stokes and Euler equations (for the most recent reviews see \cite{BT08,Const07,GBK08}). Although the Navier--Stokes equation is more relevant from the physical point of view, certain phenomena related to the non-locality of the equations are partially obscured by the presence of the viscous term and are more conveniently studied in the inviscid case. One of these phenomena is the {\it depletion} of non-linearity: generally flows behave much more  tamely than one would expect based on simple dimensional arguments and tend to suppress  singularities, approaching -- at least locally -- one of the numerous steady state solutions of the Euler equation.

Typically, inviscid incompressible flows observed in numerical simulations exhibit a much slower loss of regularity than that predicted by analytical estimates. 
However, it should be taken into account that numerical simulations aiming at studying the blow-up are mostly performed on periodic domains by employing the pseudo-spectral method with 
very smooth initial conditions, such as trigonometric polynomials. In contrast to numerical studies, analytical estimates aim at much larger classes of initial conditions for which the rate of loss of regularity may be much faster. 
Indeed, in two dimensions Bahouri and Chemin \cite{BC94} have constructed an example
of a vortex patch initial condition with exponentially fast regularity loss, thus proving that the 
corresponding estimate of the regularity loss is sharp.
%
%
Therefore, it is important to clarify how the strength of depletion may depend on the character of the initial conditions. 

One possibility to assess the strength of depletion and its dependence
on the initial conditions consists in analysing complex
singularities\footnote{ An additional reason for studying complex
singularities is that possible emergence of real singularities in
three dimensions can be detected by monitoring the dynamics of complex
singularities \cite{SSF83}, since, as has been shown by Bardos et al
\cite{BBZ76} and Benachour \cite{Ben76a,Ben76b}, a hypothetic real
singularity can only arise when a singularity located in the complex
domain hits the real domain.} of solutions of the Euler equation: for
real analytic initial conditions, solutions can be continued into the
complex domain by extending the space variables to complex
values. These solutions typicaly become singular at some complex
locations constituting, according to the numerical evidence given in
\cite{FMB03,MBF05}, a smooth manifold.  Depletion is especially
pronounced in the neighbourhood of the complex singularities: assuming
that the singular manifold has co-dimension one, it follows from the
incompressibility condition that non-linearity vanishes to the leading
order in the direction normal to it.

It has been found in \cite{PMFB06} (in the following abbreviated as
PMFB) by high-precision numerical simulations that this proliferation
of depletion in the neighbourhood of singularities entails a
significant dependence of the analytic properties of solutions on the
initial conditions: it affects the geometry but also the type of the
singularities. The singularities are non-universal: strong numerical
evidence has been given in PMFB that in the neighbourhood of the
singularities the vorticity behaves like $s^{-\beta } $ (here $s$ is
the distance to the singular manifold) with the exponent $\beta $
depending on the choice of the initial condition.

Let us now briefly describe some of the findings in PMFB. In that
work, solutions corresponding to the initial conditions $\psi _0 (z_1
, z_2 ) = \hat{F} (\twop ) \cos \twop \cdot {\bm z} + \hat{F} (\twoq )
\cos \twoq \cdot {\bm z} $ were considered in the short-time
asymptotic r\'egime.  The exponent $\beta $ was determined indirectly
by analysing the asymptotic behaviour of the Fourier coefficients of
the stream function along so-called diagonals (i.e. directions with
rational slope) in the Fourier space at high
wavenumbers. Asymptotically, the Fourier coefficients were found to
behave as product of an exponential (which gives the leading order
contribution) with an algebraic prefactor $\vert {\bf k} \vert
^{-\alpha } $ (being the first subdominant contribution). The exponent
$\alpha $ is related to the rate of divergence of the vorticity $\beta
$ via $\alpha + \beta = 7/2$ and can be determined numerically.  The
precision in determining the numerical value of the exponent $\alpha $
was sufficient to insure that the type of the singularities changes
depending on the initial conditions: two different initial conditions
were found with $\alpha = 2.66 \pm 0.01$ and $\alpha = 2.54 \pm 0.01$.
It was also found that the numerical value of the algebraic prefactor
exponent $\alpha $ depends on the angle between the vectors $\twop $
and $\twoq $, but does not depend on their lengths.

To achieve a better understanding of the non-universality of the type
of complex singularities and of their dependence on the initial
conditions a systematic study of the numerical values of $\alpha $ and
their dependence on the angle between $\twop $ and $\twoq $ is needed.
However, such a study requires a more precise determination of the
values of the exponent $\alpha $ as a function of the initial data,
since for small values of the angle between $\twop $ and $\twoq $ the
changes in the numerical value of $\alpha $ are of the same order as
the errors in $\alpha $.

Unfortunately, for the cases which have been presented in PMFB a more
precise determination of the exponent is not feasible with the
resolution achieved there.  The main reason for this is the presence
of strong intermediate asymptotics and our lack of knowledge of the
asymptotic expansion of the Fourier coefficients along diagonals in
the Fourier space.

Here we shall show that there exist initial conditions for which it is
possible to achieve high precision for the numerical value of $\alpha
$ and to determine the subdominant contributions: in the one case we
find $\alpha = 5/2$, in the other $\alpha = 3$. In Section~\ref{s:two}
we introduced a rescaled form of the two-dimensional Euler equation,
describe Taylor expansions of its solutions, their relation to complex
singularities and to the short-time behaviour of solutions. In
Section~\ref{s:shorttime} we give a detailed account of the short-time
asymptotic r\'egime, in particular with respect to the dependence on
the angle between the vectors $\twop $ and $\twoq $. In
Section~\ref{s:parallel} we study the case of initial conditions with
the angle between $\twop $ and $\twoq $ going to zero. We show that
the numerical value and the asymptotic expansion of the Fourier
coefficients can be determined very accurately. In
Section~\ref{s:antiparallel} we study the limiting case of the vectors
$\twop $ and $\twoq $ becoming anti-parallel. We show that in contrary
to the cases studied previously in PMFB the complex singularities are
not located ``above'' a symmetry point. We also determine accurately
the value of $\alpha $ and the asymptotic expansion of the Fourier
coefficients. In Section~\ref{s:general} we make some remarks on the
general case, when the angle between $\twop $ and $\twoq $ is not
restricted to any special value.  Finally, we conclude with
Section~\ref{s:conclusions}.

\section{Rescaled two-dimensional Euler equation}
\label{s:two}

\subsection{Basic equation}
\label{s:basicequation}

We begin by considering the two-dimensional Euler equation
\begin{equation}
\partial _t \Delta \psi = J (\psi , \Delta \psi ) ,
\end{equation}
on the periodic domain $\mathbb{T} _{\mathbb{R} }^2 = [0,2 \pi ) \times [0 , 2 \pi ) $. 
We chose the simplest periodic initial conditions with nontrivial dynamics, that is,  trigonometric polynomials with only two harmonics 
\begin{equation}
\psi _0 (z_1 , z_2 ) = 
\hat{F} (\twop ) e^{-i \twop \cdot {\bm z} } + 
\hat{F} (\twoq ) e^{-i \twoq \cdot {\bm z} } +  c.c.
\end{equation}
Note that after a real translation we can assume that the initial
amplitudes are real. Therefore the initial condition becomes
\begin{equation}
\label{e:init2}
\psi _0 (z_1 , z_2 ) = 
2\, \hat{F} (\twop ) \cos    \twop \cdot {\bm z}  + 
2\, \hat{F} (\twoq ) \cos    \twoq \cdot {\bm z}  .
\end{equation} 
Since we shall restrict ourselves to the initial conditions \eqref{e:init2}, we introduce 
the auxiliary coordinates
$\overline{z} _1 = \twop \cdot {\bm z} $, $\overline{z} _2 = \twoq \cdot {\bm z} $ and a rescaled time
$\tau = (\twop \wedge \twoq ) t $. Then, writing
the solution as
\begin{equation}
\label{e:rescaledEuler}
\psi (z_1 , z_2 ; t ) =  \overline{\psi }  \bigl( \overline{z} _1 , \overline{z} _2 ; (\twop \wedge \twoq ) \, t
\bigr) ,
\end{equation}
we obtain the Euler equation in the following form
\begin{equation}
\label{e:Eulerequation2}
\partial _{\tau } \overline{\Delta } _{(\twop , \, \twoq )} \overline{\psi } = \overline{J} (\overline{\psi } , 
\overline{\Delta } _{(\twop , \, \twoq )} \overline{\psi } ) ,
\end{equation}
with initial condition
\begin{equation}
\label{e:rescaledinit1}
\overline{\psi } _0 (\overline{z} _1 , \overline{z} _2 )  = 2 \, \hat{F} (\twop ) \cos \overline{z} _1 + 
2 \, \hat{F} (\twoq ) \cos \overline{z} _2 ,
\end{equation}
where $ \overline{\Delta } _{(\twop , \, \twoq )} $ denotes the Lapacian
\begin{equation}
\overline{\Delta } _{(\twop , \, \twoq )} = \vert \twop \vert ^2 \, \overline{\partial } _1^2 + 
2 \, \twop \cdot \twoq \, \overline{\partial } _1 \overline{\partial } _2 + 
\vert \twoq \vert ^2 \, \overline{\partial } _2^2 .
\end{equation}
Note that in this formulation the vectors $\twop $ and $\twoq $ do not necessarily have 
to be vectors in $\mathbb{Z}^2 $, but can take any arbitrary values in $\mathbb{R} ^2 $. 

It is obvious that if $\overline{\psi } (\overline{z} _1 , \overline{z} _2 ; t )$ is a solution of 
\eqref{e:rescaledEuler} with intitial condition $\overline{\psi } _0 $, then 
$\overline{\psi } _{\lambda } (\overline{z} _1 ,
\overline{z} _2 ; \tau _{\lambda } ) = \lambda \overline{\psi } _{\lambda } (\overline{z} _1 , \overline{z} _2 ; 
\lambda \tau ) $ is the solution of the same equation with rescaled time $\tau _{\lambda } = \lambda 
\tau $ and the initial condition $\lambda \overline{\psi } _0 $. 
Thus, the only relevant 
parameters  characterising the problem are the ratio 
$\lambda = \hat{F} (\twoq ) / \hat{F} (\twop ) $ of the two initial amplitudes $\hat{F} (\twop ) $ and $\hat{F} (\twoq ) $ and  the angle $\phi $ between the vectors $\twop $ and 
$\twoq $ together with the ratio  of lenghts 
$\eta = \vert \twoq \vert / \vert \twop \vert $ of these two vectors. 
The latter fact is immediately seen by rewriting the Laplacian as 
\begin{equation}
\overline{\Delta } _{(\twop , \, \twoq )} = \vert \twop \vert \, \overline{\Delta } _{(\phi , \, \eta )} ,
\end{equation}
and obtaining
\begin{equation}
\label{e:Eulerequation3}
\partial _{\tau } \overline{\Delta } _{(\phi , \, \eta )} \overline{\psi } = \overline{J} (\overline{\psi } , 
\overline{\Delta } _{(\phi , \, \eta )} \overline{\psi } ) ,
\end{equation}
where
\begin{equation}
\label{e:Laplacian2}
 \overline{\Delta } _{(\phi , \, \eta )} = \frac{1}{\eta } \, \overline{\partial } _1^2 + 
2 \, \cos \phi \, \overline{\partial } _1 \overline{\partial } _2 + 
\eta \, \overline{\partial } _2^2 .
\end{equation}

\subsection{Perturbative approach}
\label{s:perturb1}

From \eqref{e:Laplacian2} it follows immediately that in the case
$\eta = 1$ Equation \eqref{e:Eulerequation3} has a stationary
solution. Since, as we have mentioned, the parameter $\eta $ can in
principle take any values, for values of $\eta $ close to $1$ we try a
perturbative expansion in $\varepsilon = \eta - 1$.  The leading order
$\overline{\psi } _0 $ coincides with the initial condition
\begin{equation}
\overline{\psi } _0 (z_1 , z_2 )   = 2 \, \hat{F} (\twop ) \cos \overline{z} _1 + 
2 \, \hat{F} (\twoq ) \cos \overline{z} _2 ,
\end{equation}
whereas the next order $\overline{\psi }  _1 $ is the solution of the linearised Euler equation with a source term
\begin{equation}
\begin{split}
\partial _t \Delta _{\phi } \overline{\psi }_1 = 
& \ J (\overline{\psi } _0 , \Delta _{\phi } \overline{\psi } _1 )
+ J (\overline{\psi } _1 , \Delta _{\phi } \overline{\psi } _0 ) + 
\\
&
\ 2 \, \hat{F} (\twop ) \, \hat{F} (\twoq ) \, 
\sin z_1 \, \sin z_2 . 
\end{split}
\end{equation}

The background flow $\overline{\psi } _0 $ 
is the well-known cat's eye flow. Note that neglecting the source term and the 
perturbation term $J (\overline{\psi } _1 , \Delta _{\phi } \overline{\psi } _0 ) $ we obtain the 
shortened Euler equation which is just the passive scalar equation
\begin{equation}
\partial _t \, \Delta _{\phi } \overline{\psi } _1 + {\bm v} \cdot \nabla (\Delta _{\phi } 
\overline{\psi } _1 ) = 0,
\end{equation}
where the velocity field is determined by ${\bm v} = (\overline{\partial } _2 \overline{\psi } _0 , - 
\overline{\partial } _1 \overline{\psi } _0  ) $.
Singularities of advected fields such as $\Delta _{\phi } \overline{\psi } _1 $ have been studied in \cite{TM05a,TM05b}.
Since these singularities are determined by the singularities of the back-to-labels map and the flow is integrable, the singularities can be determined completely, see  \cite{TM05a} for the case 
$\lambda = 1$ 
($\hat{F} (\twop ) = \hat{F} (\twoq )$) and \cite{TM05b} for the more general 
case $0 \leq \lambda \leq 1$.

Coming back to the linearised Euler equation we note that it has been
extensively studied for various types of background flows. In
particular, in \cite{LLS04} flows fo Kolmogorov's type have been
studied (this type of flows, however, does not have any complex
singularities).  However, an analytic study of singularities of
solutions of the linearised time-dependent Euler equation has not been
attempted here, but certainly merits a more detailed consideration.

\subsection{Taylor series expansion} 
\label{s:Taylor}

To calculate the solutions of Equation \eqref{e:Eulerequation3} we use
the approach based on Taylor series expansion \cite{TG37}:
representing the solution as Taylor series in time
\begin{equation}
\overline{\psi } (\overline{\bm z} , t) = \sum_{n=0}^{\infty } \overline{\psi } ^{(n)} (\overline{z} ) \tau ^n ,
\end{equation}
it is obvious that the functions $\overline{\psi } ^{(n)} (\overline{z} ) $ satisfy the following recursion relation
\begin{equation}
\label{e:recursion1}
\overline{\Delta } _{(\twop , \twoq )} \overline{\psi } ^{(n+1)} = \frac{1}{n+1} \sum_{m=0}^n 
\overline{J} (\overline{\psi } ^{(m)} , \overline{\Delta } _{(\twop , \twoq )} \overline{\psi } ^{(n-m)} ) .
\end{equation}
For example, for $\overline{\psi } ^{(1)} $ we obtain
\begin{equation}
\begin{split}
\overline{\psi } ^{(1)} (\overline{z} ) = 
& 
\  2 \, \hat{F} (\twop ) \hat{F} (\twoq ) \frac{\vert \twoq \vert ^2 - 
\vert \twop \vert ^2 }{ \vert \twop - \twoq \vert ^2 } \cos (\overline{z} _1 - \overline{z} _2 ) - 
\\
&
\  2 \, \hat{F} (\twop ) \hat{F} (\twoq ) \frac{\vert \twoq \vert ^2 - 
\vert \twop \vert ^2 }{ \vert \twop + \twoq \vert ^2 } \cos (\overline{z} _1 + \overline{z} _2 ) .
\end{split}
\end{equation}
In general the functions $\overline{\psi } ^{(n)} (\overline{z} _1 , \overline{z} _2 ) $ will have the form
\begin{equation}
\label{e:Gn1}
\begin{split}
&
\overline{\psi }
^{(n)} (\overline{z} _1 , \overline{z} _2 ) = \sum_{m=0}^{2m <  n} \sum_{\sigma = 1}^{n - 2m}
2 \bigl( \hat{F} (\twop ) \bigr)^{n+1- 2m - \sigma }  \bigl( \hat{F} (\twoq ) \bigr)^{2 m + \sigma }  
\\
&
\Bigl\{  F^{(n,m)}_{\! +\ \, \sigma} 
\cos \bigl( (n+1-2m-\sigma ) \overline{z} _1 + \sigma 
\overline{z} _2 \bigr) + 
\\
&
F^{(n,m)}_{\! -\ \, \sigma}   
\cos \bigl( (n+1-2m-\sigma ) \overline{z} _1 - 
\sigma  \overline{z} _2 \bigr) \Bigr\} + 
\\
&
\sum_{m=1}^{2m \leq n }  
2 \bigl( \hat{F} (\twop ) \bigr)^{n+1- 2m}  \bigl( \hat{F} (\twoq ) \bigr)^{2 m} 
F_1^{(n,m)}  \cos \bigl( (n+1-2m) \overline{z} _1 \bigr) + 
\\
&
\sum_{m=1}^{2m \leq n }  
2  \bigl( \hat{F} (\twop ) \bigr)^{2 m}  \bigl( \hat{F} (\twoq ) \bigr)^{n+1- 2m} 
F_2^{(n,m)}  \cos \bigl( (n+1-2m) \overline{z} _2 \bigr) .
\end{split}
\end{equation} 
The coefficients $F^{(n,m)}_{\! +\ \, \sigma} $, $F^{(n,m)}_{\! -\ \, \sigma} $, $F_1^{(n,m)} $
and $F_2^{(n,m)} $ depend only on the initial vectors $\twop $ and $\twoq $, more precisely on 
the parameters $\phi $ and $\eta $. For example, for $\overline{\psi } ^{(2)} $ we have
\begin{equation}
\label{e:coeffs2}
\begin{split}
&
F^{(2,0)}_{\! +\ \, 1} =   \frac{1}{2} \, \frac{\vert \twoq \vert ^2 - \vert \twop \vert ^2 }{\vert \twop + \twoq \vert ^2 }
\, \frac{\vert \twop + \twoq \vert ^2 - \vert \twop \vert ^2 }{\vert 2 \twop + \twoq \vert ^2 } , 
\\
&
F^{(2,0)}_{\! -\ \, 1} =   \frac{1}{2} \, \frac{\vert \twoq \vert ^2 - \vert \twop \vert ^2 }{\vert \twop - \twoq \vert ^2 }
\, \frac{\vert \twop - \twoq \vert ^2 - \vert \twop \vert ^2 }{\vert 2 \twop - \twoq \vert ^2 } , 
\\
&
F^{(2,0)}_{\! +\ \, 2} =   - \frac{1}{2} \, \frac{\vert \twoq \vert ^2 - \vert \twop \vert ^2 }{\vert \twop + \twoq \vert ^2 }
\, \frac{\vert \twop + \twoq \vert ^2 - \vert \twoq \vert ^2 }{\vert \twop + 2 \twoq \vert ^2 } , 
\\
&
F^{(2,0)}_{\! -\ \, 2} =  - \frac{1}{2} \, \frac{\vert \twoq \vert ^2 - \vert \twop \vert ^2 }{\vert \twop - \twoq \vert ^2 }
\, \frac{\vert \twop - \twoq \vert ^2 - \vert \twoq \vert ^2 }{\vert \twop - 2 \twoq \vert ^2 } ,
\\
&
F_1^{(2,1)} = \frac{1}{2} \frac{ \vert \twoq \vert ^2 - \vert \twop \vert ^2 }{\vert \twop \vert ^2 }  \left( 
2 - \frac{\vert \twoq \vert ^2 }{\vert \twop + \twoq \vert ^2 }  -  \frac{\vert \twoq \vert ^2 }{\vert \twop - \twoq \vert ^2 }
\right) , 
\\
& 
F_2^{(2,1)} = - \frac{1}{2} \frac{ \vert \twoq \vert ^2 - \vert \twop \vert ^2 }{\vert \twoq \vert ^2 }  \left( 
2 - \frac{\vert \twop \vert ^2 }{\vert \twop + \twoq \vert ^2 }  -  \frac{\vert \twop \vert ^2 }{\vert \twop - \twoq \vert ^2 }
\right) . 
\end{split}
\end{equation}
Inserting the representation \eqref{e:Gn1} into Equation
\eqref{e:recursion1} we obtain a system of recursion relations for the coefficients. Since the resutling expressions turn out to be quite bulky we prefer not to list them here (see, however, \cite{WPthese07}
in a slightly different notation).

Note that contrary to solutions of the original Euler equation the
higher-order terms $\overline{\psi } ^{(n)} (\overline{z} _1 ,
\overline{z} _2 )$ do not vanish trivially for $\phi \to 0 $ (or $\phi
\to \pi $) if we take the limit keeping the ratio $\eta $ of lengths
of $\twop $ and $\twoq $ fixed. However, it is not obvious that the
limit $\phi \to 0 $ (or $\phi \to \pi $) is well-defined. Indeed, if
we try to evaluate expressions \eqref{e:coeffs2} by putting directly,
say $\phi = 0$ and $\eta = 1$ or $\eta = 2$, we obtain ill-defined
expressions because of vanishing denominators.  Thus, we have to be
careful about specifying the limit $\phi \to 0 $. Actually we are
considering a double limit: the limit of the initial modes becoming
parallel and the limit of long times, as can be seen from
\eqref{e:rescaledEuler}.  We shall come back to this question in
Section~\ref{s:shorttime}.

Let us finally make a small remark on the utility of the time Taylor
series expansion: introducing the auxiliary coordinates $\xi _1 = e^{-
i \overline{z} _1 } $ and $\xi _2 = e^{- i \overline{z} _2 } $ we can
write the solution in form of a triple Laurent series in $\tau $,
$\xi_1 $ and $\xi _2 $. The solution obtained in this form is only
meaningful inside the domain of convergence of the corresponding
series\footnote{Unless we perform analytic continuation, which is the
subject of an ongoing work.}. The domain of convergence is determined
by the complex singularities of the function $\overline{\psi } (\xi _1
, \xi _2 ; \tau ) $: in particular, because of the regularity of
two-dimensional flows we know that for real $\overline{z} _1 $,
$\overline{z} _2 $ there are no singularities for real values of $\tau
$. However, there will typically be singularities for some
complex-valued $\tau ^{\star } $, which will prevent the series from
convergence even for real-valued $\tau $'s, when $\tau > \vert \tau
^{\star } \vert $, see e.g. \cite{MOMMF81}.  Thus, the power series
representation is not well-suited for studying the temporal dynamics
of the solution. Nevertheless, it lends itself very well for analysing
space-time structure of the singularities.  Indeed, it is known that
from the asymptotic behaviour of the coefficients of a series
conclusions can be made concerning the geometry and the type of the
complex singularities of $\overline{\psi } (\xi _1 , \xi _2 ; \tau )
$, see \cite{PMFB06,WPthese07}.

\subsection{Short-time asymptotics}
\label{s:ssa}

The calculation of the solution is greatly simplified if we consider
the asymptotic short-time behaviour as in
\cite{FMB03,MBF05,PMFB06}. Indeed, at short times the Fourier
coefficients corresponding to each $\cos (k_1 \overline{z} _1 + k_2
\overline{z} _2 ) $ can be represented as
\begin{equation}
\tau ^{k_1 + k_2 - 1} 
2 \bigl( \hat{F} (\twop ) \bigr)^{k_1 }  \bigl( \hat{F} (\twoq ) \bigr)^{k_2 } 
F^{(k_1 + k_2 - 1,0)}_{\! +\ \, k_2} (1 +  O(\tau ^2 ) ) , 
\end{equation}
for $k_2 > 0 $ and 
\begin{equation}
\tau ^{k_1 - k_2 - 1} 
2 \bigl( \hat{F} (\twop ) \bigr)^{k_1 }  \bigl( \hat{F} (\twoq ) \bigr)^{-k_2 } 
F^{(k_1 - k_2 - 1,0)}_{\! -\ \, - k_2} (1 +  O(\tau ^2 ) ) , 
\end{equation}
for $k_2 < 0 $. The wave vector $k_1 $ can be assumed positive without any loss of generality.
Then, in the asymptotic r\'egime $t \to 0$ the solution will have the form
\begin{equation}
\begin{split}
&
\overline{\psi } _{\mathrm{as} } (\overline{z} _1 , \overline{z} _2 ) = 
\sum_{n=0}^{\infty }  \sum_{\sigma = 1}^n 2 \bigl( \hat{F} (\twop ) \bigr)^{n+1-\sigma } 
\bigl( \hat{F} (\twoq ) \bigr)^{\sigma } \tau ^n \times  
\\
&
\Bigl\{ 
F^{(n,0)}_{\! +\ \, \sigma}
\cos \bigl( (n+1-\sigma ) \overline{z} _1 + \sigma \overline{z} _2 \bigr) + 
\\
&
F^{(n,0)}_{\! -\ \, \sigma}  \cos \bigl( (n+1-\sigma ) \overline{z} _1 - 
\sigma  \overline{z} _2 \bigr) 
\Bigr\}
\end{split}
\end{equation}
Representing $\cos \bigl( (n + 1-\sigma ) \overline{z} _1 + \sigma \overline{z} _2 $
and  $\cos \bigl( (n + 1-\sigma ) \overline{z} _1 - \sigma \overline{z} _2 $ as a sum of two exponentials
we can write the short-time asymptotic solution as
\begin{equation}
\begin{split}
&
\overline{\psi } _{\mathrm{as} } (\overline{z} _1 , \overline{z} _2 ) = \frac{1}{\tau } 
\overline{\psi } _{+} 
(\overline{z} _1 + i \ln \tau , \overline{z} _2 + i \ln \tau ) +  
\\
&
\frac{1}{\tau } \overline{\psi } _{-} (\overline{z} _1 +  i \ln \tau , \overline{z} _2 - i \ln \tau ) + c.c. ,
\end{split}
\end{equation}
where 
\begin{equation}
\begin{split}
&
\overline{\psi } _{+} (\tilde{z} _1 , \tilde{z} _2 ) = 
\\
&
\sum_{n=0}^{\infty }  \sum_{\sigma = 1}^n \bigl( \hat{F} (\twop ) \bigr)^{n+1-\sigma } 
\bigl( \hat{F} (\twoq ) \bigr)^{\sigma }   
F^{(n,0)}_{\! +\ \, \sigma}
e^{ - i [ (n+1-\sigma ) \tilde{z} _1 + \sigma \tilde{z} _2 ] }  
\end{split}
\end{equation}
and, similarly 
\begin{equation}
\begin{split}
&
\overline{\psi } _{-} (\tilde{z} _1 , \tilde{z} _2 ) = 
\\
&
\sum_{n=0}^{\infty }  \sum_{\sigma = 1}^n \bigl( \hat{F} (\twop ) \bigr)^{n+1-\sigma } 
\bigl( \hat{F} (\twoq ) \bigr)^{\sigma }   
F^{(n,0)}_{\! -\ \, \sigma}
e^{ - i [ (n+1-\sigma ) \tilde{z} _1 - \sigma \tilde{z} _2 ] }   .
\end{split}
\end{equation}
As has been noted in \cite{PMFB06}, the functions 
\begin{equation*}
\frac{1}{\tau } \, \overline{\psi } _{+} (\overline{z} _1 + 
i \ln \tau , \overline{z} _2 + i \ln \tau ) , \ 
\frac{1}{\tau } \, \overline{\psi } _{-} (\overline{z} _1 + 
i \ln \tau  , \overline{z} _2 - i \ln \tau )
\end{equation*}
 are actually solutions of the Euler equation \eqref{e:rescaledEuler}
corresponding to initial conditions 
\begin{equation}
\label{e:simplifedinit1}
\hat{F} (\twop ) e^{-i z_1} + \hat{F} (\twoq ) e^{-i z_2 } 
\end{equation}
 and 
$\hat{F} (\twop ) e^{-i z_1} + \hat{F} (\twoq ) e^{i z_2 } $. Thus, at short times, the solution of the 
Euler equation \eqref{e:rescaledEuler} with initial condition \eqref{e:rescaledinit1} behaves as
a linear superposition of four solutions with very simple two-mode initial conditions of the type
\eqref{e:simplifedinit1}

\begin{figure}[h]
\centerline{%
\includegraphics[scale=0.25]{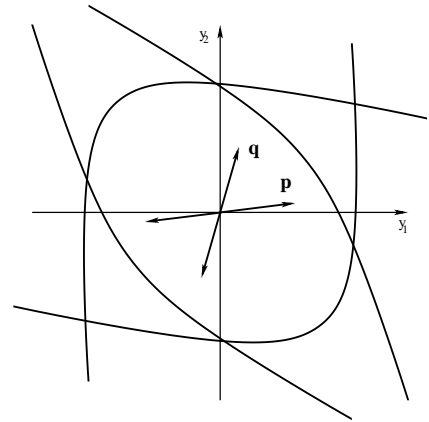}%
}
\caption{
Schematic representation of the projection of the singular manifold onto the $(y_1 ,
y_2 )$-plane at short-times. Only the parts of the singular manifold closest to the real 
domain are sketched. The singularities are represented in non-rescaled coordinates. 
Note that the vectors $\twop $ and $\twoq $ determine the asymptotic directions of 
the various branches.
}
\label{f:fourparts}
\end{figure}

Geometrically, the above observation can be explained as follows (see
Fig.~\ref{f:fourparts}): the singular manifold consists of four
branches which at short-times are located very far away from each
other.  Therefore, at short times every branch behaves as if it would
not see the presence of other branches; the nonlinear interactions
between separate branches are weak.

\section{Solutions of the Euler equation in the short-time asymptotic r\'egime}
\label{s:shorttime}

\subsection{Fourier space and pseudo-hydrodynamic formulations}
\label{s:pseudoh}

In this section we study short-time asymptotic solutions of the
two-dimensional Euler equation \eqref{e:rescaledEuler}.  Since the
coefficients $F^{(n,0)}_{\! -\ \, \sigma} $ can be determined using
the formula
\begin{equation}
F^{(n,0)}_{\! -\ \, \sigma}  (\twop , \twoq ) = (-1)^n F^{(n,0)}_{\! +\ \, \sigma}  (\twop , - \twoq ) ,
\end{equation}
we concentrate on the solution given by $ \overline{\psi } _{+} $. Here we assume that $\phi \neq 0 $. 
In what follows we put for brevity $F^{(n,0)}_{\! +\ \, \sigma}  = G^{(n)}_{\sigma } $.
The coefficients $ G^{(n)}_{\sigma } $ can be calculated by means of the following recursion relation
\begin{equation}
\label{e:recursion2}
\begin{split}
&
G^{(n)}_{\sigma } = - \frac{1}{n} \, \frac{1}{\vert (n+1-\sigma ) \twop + \sigma \twoq \vert ^2 } \, 
\sum_{m=0}^{n-1} \sum_{\tau = 0}^{\sigma } 
\\
&
\vert ((n-m)-(\sigma - \tau ) ) \twop + 
(\sigma - \tau ) \twoq \vert ^2 \, 
\\
&
[ (m+1)(\sigma - \tau ) - (n-m) \tau ] G^{(m)}_{\tau }  
G^{(n-m-1)}_{\sigma - \tau } 
\end{split}
\end{equation}
with initial values $G^{(0)}_0 = 1 = G^{(0)}_1 $. It holds\footnote{Easily seen by permuting 
$k_1 = n + 1 - \sigma $ and 
$k_2 = \sigma $ and using formula \eqref{e:recursion1}.}
\begin{equation}
G^{(n)}_{n+1-\sigma } (\twop , \twoq ) = (-1)^n G^{(n)}_{\sigma } (\twoq , \twop )
\end{equation}
Similarly to what has been noted in PMFB06, the coefficients  $ G^{(n)}_1 $ can be obtained 
explicitly
\begin{equation}
G^{(n)}_1 = (-1)^n \frac{1}{n!} 
\prod_{\sigma = 0}^{n-1} 
\frac{ \vert \sigma \twop + \twoq \vert ^2 - \vert \twop \vert ^2 }{ 
\vert (\sigma + 1) \twop + \twoq \vert ^2 }  ,
\end{equation}
as well as the coefficients $G^{(n)}_n $.

Analogously to PMFB the short-time asymptotics can be formulated as a
stationary hydrodynamical problem by introducing the
pseudo-hydrodynamic stream function
\begin{equation}
\label{e:streamf1}
G (\tilde{z} _1 , \tilde{z} _2 ) = e^{\tilde{z} _1 } + e^{\tilde{z} _2 } + 
\sum_{n=1}^{\infty } \sum_{\sigma = 1}^n G^{(n)}_{\sigma }  e^{(n+1-\sigma ) \tilde{z} _1 }
e^{\sigma \tilde{z} _2 }  .
\end{equation}
In terms of the pseudo-hydrodynamic stream function $ G (\tilde{z} _1 , \tilde{z} _2 ) $ the solution of the Euler equation \eqref{e:rescaledEuler} with  the initial condition $\hat{F} (\twop ) e^{-i z_1} + \hat{F} (\twoq ) e^{-i z_2 }  $ can be written as
\begin{equation}
\begin{split}
& 
\frac{1}{t} \, \overline{\psi } _{+} = 
\\
&
\frac{1}{t} \, G ( - i \overline{z} _1 + 
\ln t  + \ln F(\twop ) , - i \overline{z} _2 +  \ln t + \ln F(\twoq ) ) .
\end{split}
\end{equation}
This pseudo-hydrodynamic stream function solves the time independent equation 
\begin{equation}
\label{e:similarity1}
\begin{split}
&
\tilde{\Delta } _{(\twop , \twoq )} G = 
\\
&
\bigl( \tilde{\partial } _1 G + 1 \bigr)  \bigl( \tilde{\partial } _2 
\tilde{\Delta } _{(\twop , \twoq )} G \bigr)  - 
\bigl( \tilde{\partial } _2 G - 1 \bigr)  \bigl( \tilde{\partial } _1 
\tilde{\Delta } _{(\twop , \twoq )} G \bigr) .
\end{split}
\end{equation}
The initial condition is replaced by a boundary condition at infinity: for $\tilde{z} _1 \to -\infty $, 
$\tilde{z} _2 \to - \infty $ holds
\begin{equation}
G (\tilde{z} _1 , \tilde{z} _2 ) \simeq e^{\tilde{z} _1 } + e^{\tilde{z} _2 } .
\end{equation}
As has been noted in \cite{PMFB06}, Equation \eqref{e:similarity1} can be written in a form which is closer to \eqref{e:rescaledEuler} by introducing
an auxiliary function
\begin{equation}
H (\tilde{z} _1 , \tilde{z} _2 ) = G (\tilde{z} _1 , \tilde{z} _2 ) + \tilde{z} _1 - \tilde{z} _2 . 
\end{equation}
We then obtain the following pseudo-hydrodynamic equation
\begin{equation}
\tilde{\Delta } _{(\twop , \twoq )} H = \tilde{J} ( H , \tilde{\Delta } _{(\twop , \twoq )} H ) .  
\end{equation}

\subsection{Parameter-dependence of the solutions and the limiting cases $\phi \to 0$ and 
$\phi \to \pi $}
\label{s:parameterdep}

\begin{figure}[h]
\centerline{%
\includegraphics[scale=0.70]{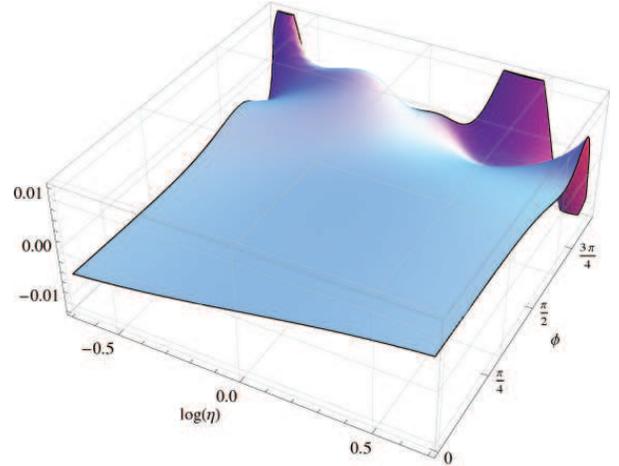}%
}
\caption{
Coefficient $G^{(7)}_{4}  (\phi , \eta ) $ in dependence on $\phi $ and $\eta $.
}
\label{f:coeffs8}
\end{figure}

To get an idea about the properties of the solutions in dependence on
the parameters we calculate the coefficients $G^{(n)}_{\sigma } (\phi
, \eta ) $ symbolically as functions of $\phi $ and $\phi $ up to
$n=8$, the coefficient $G^{(7)}_{4} (\phi , \eta ) $ being represented
on Fig.~\ref{f:coeffs8}.
From the study of the low-order coefficients we can infer that:  for (i) $\phi $ close to $0$ the 
dependence of the coefficients on both $\phi $ and $\eta $ is very smooth, 
whereas for (ii) $\phi $ close to $\pi $ the coefficients behave very irregularly.
This suggests that the case $\phi \to 0$ may have a reduced non-linearity in comparison to the case $\phi \to \pi $. With this motivation in mind we are going to study the limits $\phi \to 0$ 
and $\phi \to \pi $.

In the case $\phi \to 0 $ the short-time asymptotic solution is obtained in a straightforward fashion by putting $\phi = 0$. The Fourier coefficients of the solution can be calculated directly from the recursion relation \eqref{e:recursion1} with $\phi = 0 $. Indeed, the recursion relation \eqref{e:recursion1} 
is well-defined because for $\phi = 0 $ the Laplacian  
$(  (n+1-\sigma ) / \sqrt{\eta }  +  \sqrt{\eta } \sigma )^2 $ does not vanish. 

The case $\phi \to \pi $ is much more delicate: the values of the coefficients depend on 
the procedure we choose to calculate them.  
Consider, for example, the behaviour of the coefficient $G^{(2)}_1 (\phi , \eta ) = 
F^{(2,0)}_{\! +\ \, 1}  (\twop , \twoq ) $ 
given by \eqref{e:coeffs2}.
As function of $\phi $ and $\eta $ it is not continuous at $\eta = 1$, $\eta = 2 $ and $\phi = \pi $. Indeed, if we fix first the angle $\phi = \pi $
we obtain the expression
\begin{equation}
G^{(2)}_1 (\pi , \eta ) = 
\frac{1}{2} \, \frac{ \eta  + 1 }{\eta - 1}
\, \frac{\eta  }{\eta - 2 }  ,
\end{equation}
which  has poles\footnote{Note that if we fix a rational $\eta $, then there exist $n$ and $\sigma $
such that  the Fourier coefficient 
$ F^{(n)}_{\sigma } (\pi , \eta )  $ has a pole at $\eta $.}
at $\eta = 1 $ and $\eta = 2$. On the other hand, fixing $\eta = 1$ or 
$\eta = 2$ we will obtain finite values by considering the limits 
\begin{equation}
\lim_{\phi \to \pi } G^{(2)}_1 (\phi , 1 ) = 0 , \qquad 
\lim_{\phi \to \pi } G^{(2)}_1 (\phi , 2 ) = \frac{3}{4} .
\end{equation}
Thus, the numerical values of the coefficients for given $(\eta , \phi
) $ depend on the limiting procedure $\phi \to \pi $, $\eta ^{\prime }
\to \eta $. Since we are mostly interested in the dependence of the
solutions on the angle $\phi $, it seems natural to consider the limit
of the coefficients obtained by fixing the ratio of lengths $\eta $
first and then letting $\phi \to \pi $. The actual implementation of
this limit will be discussed in Section \ref{s:antiparallel} and
Appendix \ref{a:phieqpi}.

Note that the $\phi \to 0$ (or $\phi \to \pi$) limit obtained by fixing $\eta $ can also be 
obtained for the two-dimensional Euler equation \eqref{e:Eulerequation2}
with real-valued initial conditions \eqref{e:rescaledinit1}, see Appendix \ref{a:phieqpi}.
The relation to the so-called hydrostatic limit \cite{Bre99,Bre03,Bre08} remains to be clarified.

\section{Case of parallel initial modes}
\label{s:parallel}
%

In this section we shall consider numerical solutions for the case
$\phi = 0$. As has been found in PMFB, the value of the exponent of
the algebraic prefactor does not depend on the ratio of the lengths of
the initial vector $\eta $. In the following for convenience we shall
consider the case $\eta = 2$.  We have calculated the numerical
solution by using the recursion relation \eqref{e:recursion1} with a
precision of about $70$ significant digits and with the resolution
$N_{\mathrm{max} } = 2000$.  All numerical calculations in this and
the two following sections have been done using multiple-precision
calculation packages MPUN90 package \cite{Bay95}) and MPFR
\cite{FHLPZ07}.

We have studied asymptotic expansion of the coefficients 
$G^{(n)} (\sigma ) = G (k_1 , k_2 ) $ (where $k_1 = n+1 - \sigma $ and
$k_2 = \sigma $)  along half-lines 
\begin{equation*}
\vert {\bm k} \vert (\cos \theta , \sin \theta ) = (k_1 , k_2 ) = m (p,q)  
\end{equation*}
in the Fourier space (later we shall refer to them as {\it rational
directions}) for $\vert {\bm k} \to \infty $.  For this purpose we
have used the asymptotic interpolation procedure of van der Hoeven
\cite{vdH06}, see also \cite{PF07}. In Appendix \ref{a:asymptinter} we
briefly remind the reader the main principles of numerical analysis of
asymptotic expansions.

\begin{figure}[h]
\centerline{%
\includegraphics[scale=0.70]{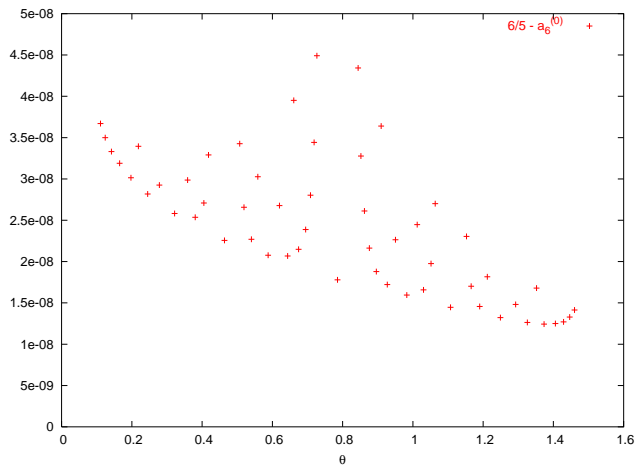}%
}
\caption{Difference between the conjectured extrapolated value
$3/\alpha = 6/5 $ for $\alpha = 5/2 $ and the numerical 
value $ 3 / \alpha _{\mathrm{num} } $ obtained 
at the sixth stage of the asymptotic 
interpolation procedure.}
\label{f:alpha}
\end{figure}

We have chosen rational directions such that $p,q < 10$, where $p$ and
$q$ are mutually prime numbers. This gives 55 different rational
directions. First, after six stages of interpolation we have
identified the leading order term and the two successive sub-leading
order terms
\begin{equation}
\label{e:asymptotics1}
G (k , \theta ) \sim  C (\theta ) 
k^{-\alpha (\theta ) } \ue ^{-\delta (\theta ) k} ,
\end{equation}
Fig.~\ref{f:alpha} shows the discrepancy of the scaling exponent $\alpha $,
determined from the interpolation value  
at the sixth stage, from the
value $5/2$ as a function of the angle $\theta $. It is seen that the numerically determined 
$\alpha _{\mathrm{num} } 
$ differs from $5/2$ by less than $10^{-7} $. We remind the reader that a theoretical argument 
has been given in PMFB suggesting that the true 
value of $\alpha $ does not depend on $\theta $.  
%
%

To confirm that the algebraic prefactor exponent $\alpha = 5/2 $ at
the next step we have tried to identify higher-order corrections to
\eqref{e:asymptotics1}. Indeed, if the higher-order corrections are
known, we will be able to estimate the parameters $\delta (\theta )$,
$\alpha$ and $C(\theta )$ more precisely. Furthermore, to obtain some
information on the analytical structure of the solutions it is
advantageous to know their asymptotic expansion at high orders.

We remark that at the sixth stage of the interpolation procedure the numerical data are of the form
\begin{equation}
\label{e:stage6}
g^{(6)} (m) = \frac{3}{\alpha } + \omega (m) ,
\end{equation}
where $m$ is the number of the point $ m (p,q)  $ on the line with rational direction $(p,q)$. 
Continuing the procedure beyond the sixth stage we have first identified the nature of 
the remainder  $\omega (m) $ by the procedure described in the Appendix \ref{a:asymptinter}. 

The identification has been performed for the coefficients
corresponding to the most populated direction $(p,q) = (1,1)$, which
is the direction with the most pronounced asymptotic behaviour. We
have found that the correction is of the form $\omega (m) \sim 1/m^2
$. This correction corresponds to the following form of the asymptotic
expansion of the Fourier coefficients
\begin{equation}
\label{e:trans1}
\begin{split}
&
G (k , \theta ) = 
\\
&
C(\theta ) k^{- \frac{5}{2} } \, \ue ^{-\delta (\theta
) k} \left[ 1 + \frac{a_1 (\theta ) }{k} + \frac{a_2 (\theta ) \ln k
}{k^2 } + O\left( \frac{1}{k^2 } \right) \right] .
\end{split}
\end{equation}
We note that the same asymptotic expansion is found for solutions of the linearised Euler 
equation considered in the short-time r\'egime, see Appendix \ref{s:perturb2}.

\begin{figure}[h]
\centerline{%
\includegraphics[scale=0.70]{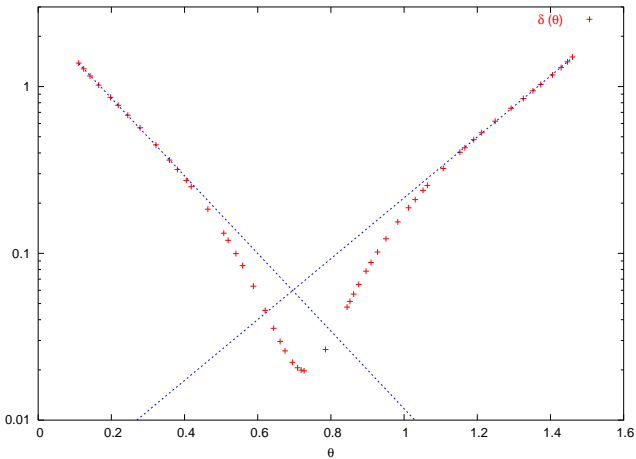}%
}
\caption{ Leading-order asymptotics: $\delta (\theta ) $ in lin-log
coordinates.  }
\label{f:delta}
\end{figure}

\begin{figure}[h]
\centerline{%
\includegraphics[scale=0.70]{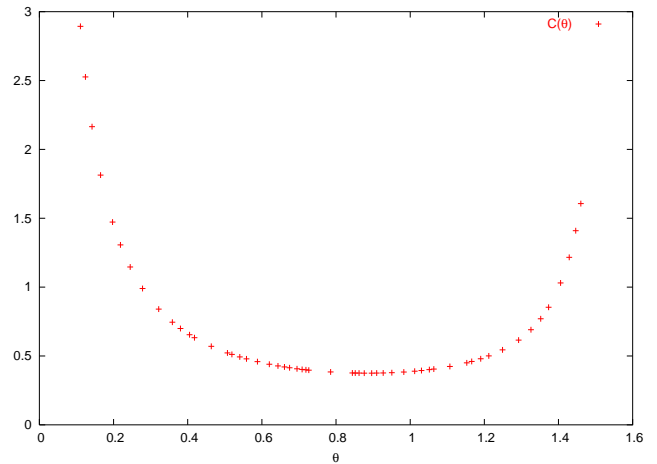}%
}
\caption{
Second subleading term $C(\theta )$. 
}
\label{f:C}
\end{figure}

It is very unlikely that the functional form of the asymptotic
expansion depends on the direction $(p,q)$ in the Fourier
space. However, for large $p$ and $q$ the number of points on the
half-lines with the rational direction $(p,q) $ is not very high, and
therefore the asymptoic behaviour may not be attained. Therefore the
values of the coefficient $a_2 (\theta ) $ that we have measured using
the asymptotic interpolation at thirteenth stage exhibit a quite large
scatter.  Due to this, for determining $\delta (\theta ) $ and
$C(\theta )$ (represented on Figs.~\ref{f:delta} and \ref{f:C}) we
have taken into account only the correction term $a_1 (\theta )/k$.


As we have stated before, averaged over all direction the absolute
error in determining $\alpha $ is of the order $10^{-7}$. However,
restricting ourselves to the most populated rational directions, we
can vastly increase the precision in determination of $\alpha $. One
possibility is for example to use one of the convergence acceleration
algorithms such as the $\rho $-algorithm for the data at sixth stage
\eqref{e:stage6}. The choice of the algorithm is fixed through the
form of the correction term being proportional to $1/m^2 $, see
\cite{Wen89}. For example, for the most populated direction $(1,1)$
the acceleration algorithm stabilises at $5/2$ with an error of the
order $10^{-16} $ after 50 iterations.  
%




%

\section{Case of anti-parallel initial modes}
\label{s:antiparallel}

In this Section we study numerically the case of antiparallel modes
$\phi \to \pi $: as we have seen before, this limit cannot be analysed
directly by taking $\phi = \pi $. Instead, we fix $\eta $, taking the
limit $\phi \to \pi $ of the recursion relation \eqref{e:recursion1}.

Here we shall only consider the case\footnote{It is not difficult to
extend the calculation to other values of $\eta $, for example $\eta =
3$ or $\eta = 4$.} $\eta = 2 $: the recursion relation is undefined
for $n+1-\sigma = 2 \sigma $ which we have to analyse separately. It
turns out that both the sum on the right-hand side of
\eqref{e:recursion1} and the Laplacian $((n+1-\sigma ) - 2 \sigma )^2
$ vanish in such a way that their ratio yields a finite limit. The
detailed calculation is not difficult but somewhat cumbersome; its
details are given in Appendix \ref{a:phieqpi}.

Numerical calculation of the coefficients reveals that the properties
of the $\phi \to \pi $-limiting solution are very different from the
properties of the solutions which have been observed before, e.g.
from those of $\phi = 0$ or $\phi = \pi / 2 $. Remember that in these
cases the Fourier coefficients had alternating sings which means that
the part of the singular manifold being closest to the real domain is
located ``above'' the point $(\pi ,0 )$ in the real space.

\begin{figure}[h]
\includegraphics[scale=0.90]{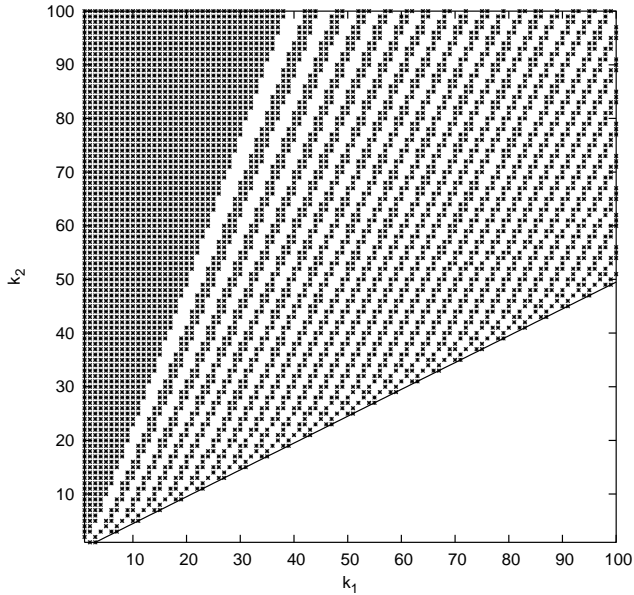}%
\caption{
Distribution of the signs of the Fourier coefficients in the in the
limiting case $\phi \to \pi $. All Fourier coefficients vanish for
$\sigma < n/3$.
}
\label{f:b1}
\end{figure}

In the present case the situation is more complicated.  Firstly, it
turns out that the Fourier coefficients $G^{(n)} (\sigma ) $ vanish
for $\sigma < n/3$ and thus the support of the solution in the Fourier
space is bounded by the half-lines $k_1=1$ and $k_2 = (k_1 - 1)/2
$. Secondly, as can be seen on Fig.~\ref{f:b1}, in the angular
sector approximately between $k_2 = (k_1 - 1)/2 $ and $k_2 = 3 k_1 $
the coefficients can have both positive and negative signs which are
distributed in an interference pattern. And thirdly, we find that in
the sector roughly between the half-lines $k_2 = 3 k_1 $ and $k_1 = 1
$ the coefficients are strictly positive.  This means that one part of
the singular manifold is located over the point $(0,0)$, whereas the
other part is found above some curve in the real space.

The geometry of the singular manifold can be studied by a
two-dimensional generalisation of the BPH-method, discussed in
\cite{WPthese07}, which is, however, out of the scope of the present
article. Note that to study the type of the singularities it is
sufficient to consider the sector with the positive coefficients,
applying to them the asymptotic interpolation procedure used
previously.

Here again for rational directions $(p,q)$, $q/p \geq 3$ the leading
order term and the two successive sub-leading order terms are
\begin{equation}
\label{e:antipasyptotics1}
G (k , \theta ) \sim  C (\theta ) 
k^{-\alpha (\theta ) } \ue ^{-\delta (\theta ) k} ,
\end{equation}
and the data at the sixth stage of the asymptotic interpolation procedure have the form 
\begin{equation}
\label{e:antipstage6}
g^{(6)} (m) = \frac{3}{\alpha } + \omega (m) . 
\end{equation}
Continuing the interpolation procedure up to the
thirteenth stage we found that for slopes $q/p $ bigger than 
$4$ the correction can be determined very accurately and it is  of the form $1/m^4 $. 
The asymptotic expansion is just an expansion in $1/\vert {\bm k } \vert $
\begin{equation}
\label{e:antiptrans1}
\begin{split}
&
G (k , \theta ) = 
\\
&
C(\theta ) k^{- 3 } \, \ue ^{-\delta (\theta
) k} \left[ 1 + \frac{a_1 (\theta ) }{k} + \frac{a_2 (\theta ) }{k^2 } + 
\frac{a_3 (\theta ) }{k^3 } +
O\left( \frac{1}{k^4 } \right) \right] .
\end{split}
\end{equation}
To increase the precision in the determination of $\alpha $ we use the $\rho $-algorithm to accelerate the convergence. We find that the constant $3/ \alpha $ at sixth stage stabilises at the value $1$ with more than $30$ digits. 

The value $\alpha = 3 $ means that the vorticity diverges as $s^{-1/2} $ in the neighbourhood of
the singular manifold.
Note that for the two-dimensional Burgers equation with initial conditions that are given by
the velocity field determined from the stream function \eqref{e:init2}, the derivative of 
velocity field diverges at the same rate. Also, the form of the asymptotic expansion 
\eqref{e:antiptrans1} (expansion in inverse powers of $k$) is identical to the asymptotic expansion we would obtain for the Burgers equation. 
Thus, we conjecture that the complex singularities of the
solutions of the Euler equation in the case $\phi \to \pi $ are of the same type as the corresponding 
singularities of the Burgers equation.
%
%

\section{Some remarks on the general case of $0 < \phi < \pi $}
\label{s:general}

We have seen that the geometry as well as the nature of singularities
are very different in the two limiting cases $\pi \to 0$ and $\phi \to
\pi $: in the former case the singularities closest to the real domain
are located over the symmetry point $(\pi , 0) $ whereas in the latter
case they lie over the point $(0,0) $ and a certain curve in the real
domain. Furthermore, in these two cases the values of the scaling
exponents ($\alpha = 5/2 $ in the former and $\alpha = 3 $ in the
latter case) and the functional forms of the asymptotic expansions are
different.  Therefore, it is not surprising to find that the general
case $0 < \phi < \pi $ exhibits strong intermediate asymptotics,
making it difficult to determine the actual high-$k$ behaviour of the
Fourier coefficients.  In particular this impedes a precise
determination of the scaling exponent $\alpha $ and of the functional
form of the asymptotic expansion.

\begin{figure}[h]
\includegraphics[scale=0.90]{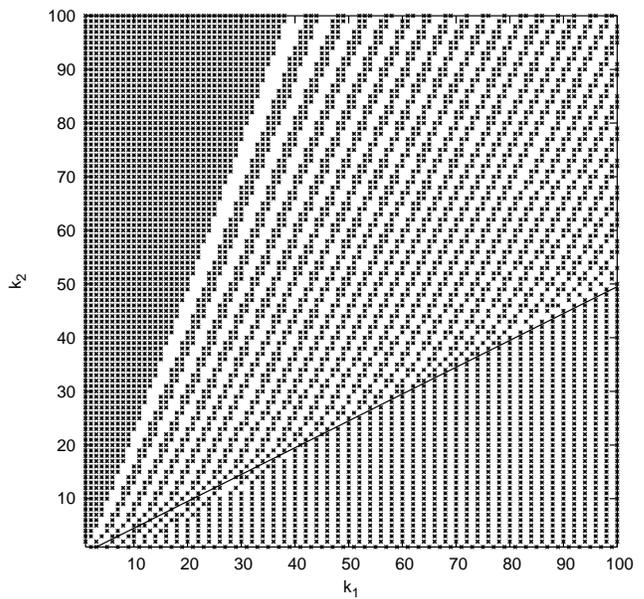}%
\caption{
Sign distribution for an angle close to $\pi $, here $\phi = 24/25 \pi $.
}
\label{f:nearantip}
\end{figure}

The easiest way to see the interplay between the $\phi \to 0 $ and
$\phi \to \pi $ limiting behaviour is to look at the case when $\phi $
is very close to $\pi $. First of all, in this case the Fourier
coefficients have non-vanishing values almost everywhere.  An
inspection of the sign distribution of the Fourier coefficients
(Fig.~\ref{f:nearantip})
shows the existence of three different regions: the first region is
roughly the one between the $k_1 $ axis and the half-line with the
slope $1/2$. The Fourier coefficients in this sector have alternating
signs. The corresponding part of the singular manifold is therefore
located over the point $( \pi , 0 )$. Note that in the case $\phi =
\pi $ the Fourier coefficients in this sector vanish. The second
region corresponds to the sector with the interference pattern in the
case $\phi = \pi $. In the third region between the half-line with the
slope $3$ and the $k_2 $-axis the Fourier coefficients are strictly
positive. The corresponding part of the singular manifold is therefore
located over the point $(0,0)$. The singular manifold has a quite
interesting geometry, however, its detailed study is out of the scope
of the present article and will be presented elsewhere.

When the angle $\phi $ is decreased from $\pi $ to $0$ the region with
alternating sings proliferates until all the Fourier coefficients
become alternating in their signs. The
threshold value of the angle is conjectured to be\footnote{This
threshold value applies only to the case $\eta = 2$. In the more
general case its value is $\arccos (-1/2 \eta) $. The threshold values
of $\phi$ have been estimated using symbolically calculated
expressions for the Fourier coefficients.}  $\phi = \arccos (-1/4) $.

In view of the complicated geometry of the singular manifold it is not
surprising that a precise determination of the scaling exponent is
very difficult and needs very high resolutions. One of the main
factors reducing the precision in the determination of $\alpha (\phi
)$ are the subdominant contributions which strongly depend on the
directions in the Fourier space in which the Fourier coefficients are
evaluated\footnote{This introduces an artificial dependence of the
scaling coefficients obtained numerically on the direction $\theta $
in the Fourier space. As we have mentioned before, it can be shown
analytically that the value of the scaling exponent does not depend on
$\theta $.}. To increase the precision one would have to take into
account these subdominant corrections.

However, with resolutions of order $N_{\mathrm{max} } = 2000$ we have
not succeeded in determining the functional form of the asymptotic
expansion of the Fourier coefficients and there could not obtain the
numerical values of the scaling exponents with a precision comparable
to the one found in the cases $\phi \to 0$ and $\phi \to \pi
$. Therefore, a systematic study of the dependence of the scaling
exponents on the angle is not feasible at the moment.

%
%
%
%
%
%
%
%
%
%
%

\section{Conclusions}
\label{s:conclusions}

In the present paper we have shown that, even for the simplest initial
conditions with non-trivial dynamics, complex singularities of the
solutions of the two-dimensional Euler equation can be of very
different nature, corresponding to the values of the scaling exponent
$\alpha = 5/2 $ and $\alpha = 3$.

This has been achieved by identifying two particularly simple cases
for which the exponents and therefore the type of the singularities
can be determined very accurately. Furthermore, we were able to
determine the subdominant terms. This increases the certainty in the
numerical value of the scaling exponents far beyond the cases which
have been analysed in PMFB.

In the limiting case $\phi \to 0$ it has been found that the value of
the scaling exponent $\alpha $ coincides with the value of the scaling
exponent for the solutions of the linearised Euler equation.
Furthermore, we have numerical evidence that the structure of the
asymptotic expansion in the Fourier space is the same in both cases.
Thus, non-linearity is suppressed in this case, which constitutes an
example of a dynamically nontrivial completely depleted flow.

Actually, the case $\phi \to 0$ can be treated by means of asymptotic
expansions near the edges, i.e. the regions for which $\theta $ is
either close to $0$ or close to $\pi $. In particular, the value
$\alpha = 5/2 $ can be found both by Maple assisted calculations and
asymptotic arguments\footnote{A.~Gilbert, private communication.}.

In  the case $\phi \to \pi $ the scaling exponent $\alpha = 3 $ is the same as for solutions of the two-dimensional Burgers equation. Also in this case the structure of the asymptotic expansion coincides in both cases. This case is an example of a flow with a very  strong nonlinearity, probably the strongest possible for the initial conditions of trigonometric polynomial type.

The difference between the two cases $\phi \to 0 $ and $\phi \to \pi $ is reflected not only in the type of the singularities:
we have found that in these two cases the geometry of complex singularities is completely different. 
In the case $\phi \to 0$ all (except one) Fourier coefficients of the solution have regularly alternating
signs which in the physical space is expressed by the fact that  the part of the singular manifold  
closest to the real domain is located ``over'' the symmetry point $(\pi , 0)$.  In the case $\phi \to 
\pi $ only a part of the Fourier coefficients has regular signs being  positive. The 
remaining coefficients do not have such regularity in the signs. The physical space counterpart
means that the part of the singular manifold closest to the real domain is located partially ``over''
the point $(0,0) $ and partially over some curve in the real domain. 

For intermediate angles $\phi $ lying between $0$ and $\pi$, for example in the SOC case $\phi = \pi / 2 $ we have not been able to determine the subdominant terms in the asymptotic expansion. Therefore, the uncertainty in the exact value of the scaling exponent, in particular in dependence on the angle $\phi $ is quite high.

What do our findings mean in consideration of the question of well-posedeness of the 
three-dimensional Euler equation? Firstly we note that the short-time asymptotic analysis of the 
kind presented here is easily extendable to three dimensions. Some preliminary results in this 
direction have been obtain in \cite{WPthese07} confirming non-universality of the nature of complex 
singularities in three dimensions  as well. Secondly one very important unresolved issue remains:
does the nature of complex singularities change when we go beyond the short-time asymptotics? 
Although some indications have been given in PMFB that the nature of singularities does not change 
we have no definitive answer yet.

We expect that the extent to which the non-linearity is depleted is strongly dependent on the 
initial conditions which we can also view as a dependence on the large-scale structures in the 
flow. We cannot exclude that even if solutions of the three-dimensional Euler equation generically 
stay regular, there exist initial conditions -- such as the Kida--Pelz -- with very strong non-linearity
which could produce a finite-time blow up.

Finally we note that the algebraic approach of the type used here can be extended to the 
viscous case. The short-time asymptotic r\'egime of the inviscid case corresponds 
in the Navier--Stokes case to the 
asymptotic r\'egime of small Reynolds numbers.

%

\begin{acknowledgments}
The author would like to thank U.~Frisch for encouragement, support and many stimulating insights. A.~Gilbert is acknowledged for his constant interest in this work. The author also thanks T.~Matsumoto, J.~Bec, D.~Mitra and D.~Vincenzi for many interesting discussions. 
\end{acknowledgments}

\vspace{3mm}

\appendix
\renewcommand{\theequation}{\thesection.\arabic{equation}}
\section{Asymptotic expansions of Fourier/Taylor coefficients}
\setcounter{equation}{0}
\label{a:asymptinter}

Let us consider a two-dimensional array $\hat{G} (k_1 , k_2 ) $ with $k_1 , k_2 \geq 0 $. 
One possibility to study the asymptotic 
behaviour of $\hat{G} (k_1 , k_2 ) $ at infinity is to proceed by simple reduction, analysing the behaviour of the one-dimensional 
arrays\footnote{The array $\hat{g} (m p , m q) $ is closely connected to the so-called diagonal of a 
double Taylor series, see \cite{STs84}.}
\begin{equation}
\hat{g} _{(p,q)} (m)  = \hat{G} (m p , m q) , 
\end{equation}
for relatively prime positive numbers $p$, $q$. An asymptotic 
expansion of $\hat{g} _{(p,q)} (m) $ for $m \to \infty $ yields then an asymptotic expansion of $\hat{G} (k_1 , k_2 ) $
in polar coordinates $(k_1 , k_2 ) = 
\vert  {\bm k} \vert ( \cos \theta , \sin \theta ) $ for $\vert {\bm k} \vert $. Note that other approaches have been suggested 
in \cite{vdH06}. They are not discussed here.

The asymptotic expansion of $\hat{g} _{(p,q)} (m) $ is found by applying the asymptotic interpolation procedure  \cite{vdH06}. 
Due to the presence of complex singularities the leading order and the two first sub-leading 
order contributions are 
\begin{equation}
\label{e:appasympt1}
\hat{g} _{(p,q) } (m) \simeq  C_{(p,q)}  m^{-\alpha } \ue ^{- \delta _{(p,q)} m } ,
\end{equation}
see also \cite{PMFB06, WPthese07}.  

The main idea of the asymptotic interpolation procedure is to find for $\hat{g} _{(p,q)} $ 
such sequence transformations -- stages of the interpolation procedure -- 
which would ``strip off'' the sequence $\hat{g} _{(p,q)} $ the leading order terms $ e^{-\delta _{(p,q)} m}
$ and $m^{-\alpha } $ without knowing the precise value of the constants $\delta _{(p,q)} $ and 
$\alpha $. One possible choice of transformations is  the one which has been given in 
\cite{PF07}:
\begin{equation}
\label{e:sixstages}
\mathrm{\bf SR}  \, , \  -  \mathrm{\bf D}  \, , \  \mathrm{\bf I}  \, , \  \mathrm{\bf D}   \, , \  
\mathrm{\bf D}  \, , \  \mathrm{\bf D}  , 
\end{equation}
where the sequence transformations are: $\mathrm{\bf I}  $ taking the inverse $\hat{g} (m) \rightarrow 1/ \hat{g} (m) $, $\mathrm{\bf SR}  $ taking the second ratio 
$\hat{g} (m) \rightarrow \hat{g} (m)  \hat{g} (m-2) / (\hat{g} (m-1) )^2 $ and 
$\mathrm{\bf D}  $ taking the difference 
$\hat{g} (m) \rightarrow \hat{g} (m) - \hat{g} (m-1) $. After applying these sequence transformations 
the leading order of the obtained sequence together with corrections will be $3/\alpha + \omega (m) $.

The main problem in studying the asymptotics of the Fourier coeffcients is the correct identification of the 
 correction term $\omega (m)$. Indeed, once its form is known we can draw conclusions about the further sub-leading order terms in the asymptotic expansion \eqref{e:appasympt1} of $\hat{g} (m) $. 
If $\omega (m) $ is to the leading order of the type $m^{-\gamma } $, as in the case of the limits $\phi \to 0$ and $\phi \to \pi $, we can identify the constant $\gamma $ by applying to the data after 
\eqref{e:sixstages} the following  transformations
\begin{equation}
 \label{e:thirteenstages}
\mathrm{\bf D}  \, , \  \mathrm{\bf I}  \, , \  \mathrm{\bf R}  \, , \  \mathrm{\bf D}   \, , \  
\mathrm{\bf I}  \, , \  \mathrm{\bf D} \, , \  \mathrm{\bf D} .
\end{equation}
The leading order term after these transformations will have the form $2 / (\gamma + 1) $. In the cases 
$\phi \to 0 $ and $\phi \to \pi $ the constant $\gamma $ can be identified as $\gamma = 2 $ and $
\gamma = 4 $ respectively.

\section{Parallel and anti-parallel modes limits of the two-dimensional Euler equation}
\label{a:phieqpi}

\subsection{Two-dimensional Euler equation in the $\phi \to 0 $ limit}

We begin by considering the two-dimensional Euler equation \eqref{e:Eulerequation3}
with initial conditions \eqref{e:rescaledinit1}.
Putting  formally $\phi = 0 $ (or $\phi = \pi $) we obtain 
\begin{equation}
\label{e:hydrostat1}
\begin{split}
&
\partial _t \bigl( \frac{1}{\sqrt{\eta } } \, \overline{\partial } _1 \pm \sqrt{\eta } \, \overline{\partial } _2 
\bigr)^2  \overline{\psi } _{\mathrm{form} }  
= 
\\
&
\overline{J}  \Bigl( \overline{\psi } _{\mathrm{form} } , 
\bigl( \frac{1}{\sqrt{\eta } }  \, \overline{\partial } _1 \pm 
\sqrt{\eta }  \, \overline{\partial } _2 \bigr)^2  \overline{\psi } _{\mathrm{form} }  \Bigr) ,
\end{split}
\end{equation}
where the initial condition remain unchanged and the 
signs $+$ and $-$ corresponds to the cases
$\phi = 0 $ and $\phi = \pi $.
 Unfortunately this equation  is ill-posed, because, 
 contrarily to the Laplacian $\Delta _{(\phi , \, \eta )} $
in the case $0 < \phi  < \pi $, the 
operator $ ( (1/\sqrt{\eta } ) \, \overline{\partial } _1 \pm \sqrt{\eta } \, \overline{\partial } _2 )^2 $ 
has a non-vanishing kernel.  

Here we show that if we first fix $\eta $ and then let $\phi \to 0 $ we obtain a well-defined
equation for the $\phi \to 0 $ limit of  \eqref{e:Eulerequation3}. 
Consider the two-dimensional Euler equation in the Fourier space
\begin{equation}
\label{e:FourierEuler}
\begin{split}
&
\partial _t \hat{\psi } (k_1 , k_2 ) = 
\\
&
- \frac{1}{\vert \twop k_1 + \twoq k_2 \vert ^2 } 
\sum_{ {\bm l}  + {\bm l} ^{\prime } = {\bm k} }  {\bm l} \wedge {\bm l} ^{\prime } 
\hat{\psi } (l_1 , l_2 )  \vert \twop l_1^{\prime } + \twoq l_2^{\prime } \vert ^2 
\hat{\psi } (l_1^{\prime } , l_2^{\prime } )   .
\end{split}
\end{equation}
Clearly, for $\twop = (1,0) $ and $\twoq = ( \eta , 0 ) $, which corresponds to the limit 
$\phi \to 0 $ the denominator on the right-hand side vanishes. However, as we shall see now,
the numerator also vanishes in such a way that the ratio of the both gives a finite number in the 
limit $\phi \to 0$.

We rearrange the sum in such a way that the term $\hat{\psi } (l_1 , l_2 ) 
\hat{\psi } (l_1^{\prime } , l_2^{\prime } ) $ appears only once for every ${\bm l}$. Thus, we obtain 
terms of the form
\begin{equation}
{\bm l} \wedge {\bm l} ^{\prime } \, \bigl( \vert \twop l_1^{\prime } + \twoq l_2^{\prime } \vert ^2 
- \vert \twop l_1 + \twoq l_2 \vert ^2 \bigr) \, 
\hat{\psi } (l_1 , l_2 )  \hat{\psi } (l_1^{\prime } , l_2^{\prime } ) , 
\end{equation}
where ${\bm l} + {\bm l} ^{\prime } = {\bm k} $. Noting that  
\begin{equation}
\begin{split}
&
\vert \twop l_1^{\prime } + \twoq l_2^{\prime } \vert ^2  - \vert \twop l_1 + \twoq l_2 \vert ^2 
=
\\
&
\Bigl( \frac{1}{\sqrt{\eta } } k_1 + \sqrt{\eta } k_2 \Bigr)^2  - 2 k_1 k_2 ( 1 - \cos \phi ) -
2 l_1 \Bigl( \frac{1}{\eta } k_1 + k_2 \Bigr)  + 
\\
&
2 l_1 k_2 (1 - \cos \phi ) - 
2 l_2 (k_1 + \eta k_2 )  + 2 l_2 k_1 (1 - \cos \phi ) ,  
\end{split}
\end{equation}
and 
\begin{equation}
\vert \twop k_1 + \twoq k_2 \vert ^2 =  \Bigl( \frac{1}{\sqrt{\eta } } k_1 + \sqrt{\eta } k_2 \Bigr)^2 
- 2 k_1 k_2 (1 - \cos \phi ) , 
\end{equation}
we see that when $ k_1 + \eta k_2 = 0 $, it follows
\begin{equation}
\frac{ \vert \twop l_1^{\prime } + \twoq l_2^{\prime } \vert ^2 
- \vert \twop l_1 + \twoq l_2 \vert ^2 }{ \vert \twop k_1 + \twoq k_2 \vert ^2 } = 
\frac{k_1 k_2 - l_1 k_2 - l_2 k_1 }{k_1 k_2 }  \neq 0 , 
\end{equation}
for $k_1 , k_2 \neq 0 $. The only case where the above remark does not apply is when 
there exists an ${\bm l} $ such that $2 {\bm l} = {\bm k} $, in which case 
we obtain
\begin{equation}
\frac{ \vert \twop l_1^{\prime } + \twoq l_2^{\prime } \vert ^2 
}{ \vert \twop k_1 + \twoq k_2 \vert ^2 } = 
\frac{1}{4}     . 
\end{equation}
Thus, the expression on the right-hand side of  \eqref{e:FourierEuler} is well-defined in the
limit $\phi \to 0$.  The analytic properties of this limiting solution and relation to solutions of 
the so-called hydrostatic equation certainly merit a  further study. We also note that a 
similar limiting solution can be constructed for the two-dimensional Navier--Stokes equation.

\subsection{Short-time recursion relation in the $\phi \to \pi $ limit}

Let us consider the limiting form of the recursion relation \eqref{e:recursion1} in the case $\eta = 2$
and $\phi \to \pi $. Firstly, we remark that in the sum on the right-hand-side of this recursion relation the
terms containing the product $G^{(m)}_{\tau }  G^{(n-m-1)}_{\sigma - \tau } $ appear twice:  firstly for 
$(m, \, \tau )$ and secondly for $(n-m-1, \, \sigma - \tau )$. Thus we can regroup the recursion relation 
in such a way as that the product $G^{(m)}_{\tau }  G^{(n-m-1)}_{\sigma - \tau } $ appears in only one 
summand, which has the form
\begin{equation}
\begin{split}
&
K_{(\twop , \twoq ) } ( n , \sigma \vert m ,\tau )  \times
\\
&
\bigl[ (m+1) (\sigma - \tau ) - (n-m) \tau \bigr]  
G^{(m)}_{\tau }  G^{(n-m-1)}_{\sigma - \tau }  , 
\end{split}
\end{equation} 
where
\begin{equation}
\begin{split}
&
K_{(\twop , \twoq ) } ( n , \sigma \vert m ,\tau ) = 
\vert  \bigl( (n-m) - (\sigma - \tau ) \bigr) \twop + (\sigma - \tau ) \twoq \vert ^2 -
\\
&
\vert (m+1 - \tau ) \twop + \tau \twoq \vert ^2 
\end{split}
\end{equation}

Secondly, since for $\eta = 2 $ the recursion relation \eqref{e:recursion1}  becomes undefined
only when $n+1 - \sigma = 2 \sigma $, we concentrate on the $\phi \to \pi $-limit  of the recursion relation 
only for $n = 3\sigma - 1$. 
As we shall see now, when $\phi \to \pi $
\begin{equation}
\begin{split}
&
\frac{1}{\vert (n+1 - \sigma ) \twop + \sigma \twoq \vert ^2 } \, 
K_{(\twop , \twoq ) } ( n , \sigma \vert m ,\tau ) =
\frac{1}{2 \sigma ^2 } 
\\
&
\Bigl\{ \bigl( (n-m) - (\sigma - \tau ) \bigr) (\sigma - \tau ) - 
(m + 1 - \tau ) \tau \Bigr\} ,
\end{split}
\end{equation} 
and has thus a finite value.  

Indeed, assume that $n_1 $, $n_1^{\prime } $, $n_2 $ and $n_2^{\prime } $ are integers such that
 $n_1 + n_1^{\prime } = 2 \sigma $ and $n_2 + n_2^{\prime } = \sigma $.
Then it holds 
\begin{equation}
\label{e:regroup1}
\vert n_1 \twop + n_2 \twoq \vert ^2 - \vert n_1^{\prime } \twop + n_2^{\prime } \twoq \vert ^2 
= 4 (n_1 n_2 - n_1^{\prime } n_2^{\prime } ) (1 + \cos \phi )
,
\end{equation}
where we have used the expressions $\twop = (1,0) $ and  $\twoq = 2 (\cos \phi , \sin \phi ) $. 
Similarly we obtain
\begin{equation}
\vert (n+1 - \sigma ) \twop + \sigma \twoq \vert ^2 = 8 \sigma ^2 (1 + \cos \phi ) ,
\end{equation}
and therefore the term $1 + \cos \phi $ is cancelled.

\section{Perturbative study of the short-time asymptotics}
\label{s:perturb2}

Here we present a simple linear model which turns out to have the same asymptotic structure as the solution of 
the short-time Euler equation in the limit $\phi = 0 $.

As we have noted in Section \ref{s:perturb1} for small values of $\varepsilon = \eta - 1$ we can 
apply a perturbative ansatz to Equation \eqref{e:similarity1}, which will give us to the first sub-leading order 
\begin{equation}
\label{e:LPDE1}		
\begin{split}
&
(1 - \ue ^{\check{z} _2 } )  \check{\partial } _1 \tilde{\Delta } 
_{(\phi , 1)} G_1 + 
(1 + \ue ^{\tilde{z} _1 } )  \tilde{\partial } _2 
\tilde{\Delta } _{(\phi , 1)} G_1 -
\tilde{\Delta } _{(\phi , 1)} G_1 + 
\\
&
\ue ^{\tilde{z} _2 } 
(\tilde{\partial } _1 G_1 ) - \ue ^{\tilde{z} _1 } (\check{\partial }
_2 G_1 ) =  - 2 \, \ue ^{\tilde{z} _1 } \ue ^{\tilde{z} _2 } .
\end{split}
\end{equation}
Neglecting the source term and the terms $ \ue ^{\tilde{z} _2 }  (\tilde{\partial } _1 G_1 ) - \ue ^{\tilde{z} _1 } (\check{\partial } _2 G_1 ) $ we obtain the passive scalar model which has been introduced 
in \cite{PMFB06}. For this model the singular manifold can be described analytically,
and furthermore, we can determine analytically the type of singularities of the field 
$\tilde{\Delta } _1 G_1 $, which turn out to be poles.  So far we have not been able to handle Equation
\eqref{e:LPDE1} satisfactorily by analytic tools. Therefore we have studied solutions of this equation by using high-precision numerics (for these calculations we have used  a precision of about 70 digits and
a resolution $N_{\mathrm{max} } = 3000$.). 

We have found that the presence (or absence) of the source term and of the term $ \ue ^{\tilde{z} _2 }  (\tilde{\partial } _1 G_1 ) - \ue ^{\tilde{z} _1 } (\check{\partial } _2 G_1 ) $ 
changes neither the location of the singularities nor the value of algebraic prefactor exponent $\alpha $. Using the asymptotic interpolation procedure  of we were able to determine the asymptotic expansion of the solution in the Fourier space, which has the same 
form as the asymptotic expansion \eqref{e:trans1} 
of the short-time Euler equation in the limit $\phi \to 0 $.

Note that the form of singular manifold and the scaling exponent $5/2 $ (or $1$ in the physical space) do not depend on the parameter $\phi $. Thus, the perturbative expansion predicts a universal exponent and thus, gives wrong predictions when $\phi \neq 0 $. Still, in the case $\phi = 0 $
it gives the correct form of the asymptotic expansion.

\end{document}